\documentclass{acm-proc-article-sp}
\usepackage{cite}
\usepackage{float}
\usepackage{hyperref}
\usepackage{url}
\usepackage{array}
\usepackage{enumitem}
\usepackage{wrapfig}
\usepackage{graphicx}
\usepackage{tabularx, longtable}
\usepackage[numbers]{natbib}
\let\oldtabular\tabular 
\renewcommand{\tabular}{\footnotesize\oldtabular}

\begin{document}
\title{SoK: Applying Machine Learning in Security - A Survey}
\numberofauthors{1} 
\author{
\alignauthor
Heju Jiang\textsuperscript{*} , Jasvir Nagra, Parvez Ahammad \titlenote{Corresponding authors} \\
       \affaddr{Instart Logic, Inc.}\\
       \email{\{hjiang, jnagra, pahammad \}@instartlogic.com} \\
}
\maketitle

\begin{abstract}
The idea of applying machine learning(ML) to solve problems in security domains is almost 3 decades old. As information and communications grow more ubiquitous and more data become available, many security risks arise as well as appetite to manage and mitigate such risks. Consequently, research on applying and designing ML algorithms and systems for security has grown fast, ranging from intrusion detection systems(IDS) and malware classification to security policy management(SPM) and information leak checking. In this paper, we systematically study the methods, algorithms, and system designs in academic publications from 2008-2015 that applied ML in security domains. 98\% of the surveyed papers appeared in the 6 highest-ranked academic security conferences and 1 conference known for pioneering ML applications in security. We examine the generalized system designs, underlying assumptions, measurements, and use cases in active research. Our examinations lead to 1) a taxonomy on ML paradigms and security domains for future exploration and exploitation, and 2) an agenda detailing open and upcoming challenges. Based on our survey, we also suggest a point of view that treats security as a game theory problem instead of a batch-trained ML problem.
\end{abstract}
%
%
%
%
%
%
\section{INTRODUCTION AND MOTIVATION}
Since Dorothy Denning's seminal 1987 paper on intrusion detection \cite{Denning:1987ids}, ML and data mining(DM) have steadily gained attention in security applications. DARPA's 1998 network intrusion detection evaluation \cite{DARPAIDS1998}, and KDD(Conference on Knowledge Discovery and Data Mining) Cup's 1999 challenge \cite{kdd1999:data, kdd1999:result} have raised profile of ML in security contexts. Yet, constrained by hardware and system resources\cite{kdd1999:result}, large-scale ML applications did not receive much attention for many years. 

In 2008, ACM Conference on Computer and Communications Security(CCS) hosted the 1st Artificial Intelligence in Security(AISec) Workshop, which has since been a dedicated venue at a top-level security conference for the intersection of ML and security. From 2008, the pace of research and publicity of ML in security started to accelerate in academic communities (section 2.3), and industry venues (e.g. Black Hat, RSA) also shifted interests. For instance, ML in security was still a topic of minority interest at Black Hat USA 2014 in August \cite{blackhat2014}, but at RSA 2016 in February, the majority of vendors claimed to deploy ML in their products \cite{RSA2016}. A part of this shift may be motivated by the sudden increase in blackswan events like the discovery of CRIME, BEAST and Heartbleed vulnerabilities.  The discovery of these vulnerabilities suggest that organizations may be attacked via previously unknown classes of attacks. To defend against these types of attacks requires monitoring not just for known vectors attacks, but also for behavior suggestive of a compromised machine. The latter requires the gathering and analysis of much larger sets of data. 

Advances in hardware and data processing capacities enabled large-scale systems. With increasing amount of data from growing numbers of information channels and devices, the analytic tools and intelligent behaviors provided by ML becomes increasingly important in security. With DARPA's Cyber Grand Challenge final contest looming \cite{DARPACGC2016}, research interest in ML and security is becoming even more conspicuous. Now is the crucial time to examine research works done in ML applications and security. To do so, we studied the state-of-art of ML research in security between 2008 and early 2016, and systematize this research area in 3 ways: 
\begin{enumerate}[nosep]
\item We survey cutting-edge research on applied ML in security, and provide a high-level overview taxonomy of ML paradigms and security domains.
\item We point to research challenges that will improve, enhance, and expand our understanding, designs, and efficacy of applying ML in security.  
\item We emphasize a position which treats security as a game theory problem. 
\end{enumerate}
\begin{figure*}
\includegraphics[height=1.5in, width=7in]{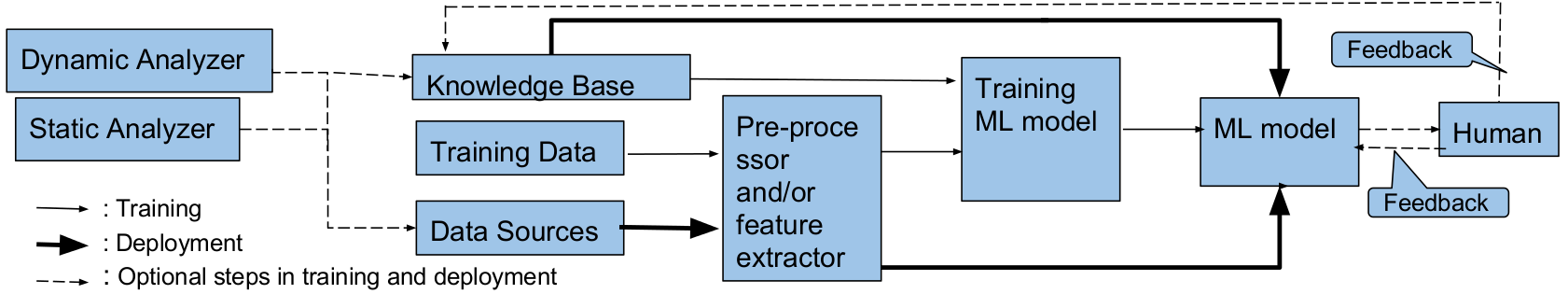}
\caption{Generalization of surveyed ML System Designs in Security; details in section 2.1}
\end{figure*}
\section{OVERVIEW: Structure \& Scope}
While we realize there are different ways to classify existing security problems based on purpose, mechanism, targeted assets, and point of flow of the attack, our SoK's section structure is based on the ``Security and Privacy'' category of 2012 ACM Computing Classification System\cite{acm2012:classification}, which is a combination of specific use cases(e.g. malware, phishing), technique (e.g. information flow), and targeted assets(e.g. web application, proxies). We present the state-of-art ML applications in security as the following: Section 3 and Table 2 \& 3 discusses Network Security\footnote{All papers are listed in chronological order with the first author's last name followed by venue acronym and year.}, Section 4 and Table 4 surveys Security Services, Section 5 and Table 5 specifies advances in Software \& Applications Security, Section 6 and Table 6 \& 7 lays out taxonomy for System Security, and Section 7 and Table 8, 9 \& 10 summarizes progress since 2008 in Malware Detection, IDS, and Social Engineering. Throughout the survey, we share our frameworks for ML system designs, assumptions, and algorithm deployments in security. 

We focus our survey on security \emph{applications} and security-related ML and AI problems on the \emph{defense} side, hence our scope excludes theories related to security such as differential privacy and privacy-preservation in ML algorithms\cite{CCS:AISec2016TOpicScope, chow2008detecting,johnson2013privacy}, and excludes ML applications in side channel attacks such as \cite{panchenko2016website, sidechannel:uncoverphrases:encryptedvoip, sidechannel:nlpeavesdrop:videosystem}. Partly because there is already a 2013 SoK on evolution of Sybil defense\cite{alvisi2013sok:sybildefense} in online social networks(OSN), and partly because we would like to leave it as a small exercise to our readers, we excluded Sybil defense schemes in OSN as well\cite{danezis2009sybilinfer, usenix2013:svmcluster:sybilosn, yu2009dsybil, xing2012scalable, ciglan2013community:graphmining}. Still with a broad base, we propose an alternative position to frame security issues, and we also recommend a taxonomy for ML applications in security use cases. Yet, we do not conclude with a terminal list of ``right'' or ``correct'' approaches or methods. We believe that the range of the applications is too wide to fit into one singular use case or analysis framework. Instead, we intend this paper as a systematic design and method overview of thinking about researching and developing ML algorithms and applications, that will guide researchers in their problem domains on an individual basis. We target our work to security researchers and practitioners, so we assume that our readers have general knowledge for key security domains and awareness of common ML algorithms, and we also define terms when needed.

\begin{table*}
\centering
\begin{tabular}{|l|l|l|l|l|l|} \hline
\multicolumn{1}{|l|}{}& {Supervised}& {Semi-supervised}& {Unsupervised}& {HITL}&{Game Theory} \\ \hline 
\multicolumn{6}{|l|}{\textbf{Attacker Type}} \\ \hline 
\multicolumn{1}{|l|}{\textit{Passive}}& {3.1/4/5/6.1/6.2/7.1/7.2/7.3}&{5/6.1/7.2}&{3.1/3.2/5/6.1/6.2/7.1/7.2}&{5/6.1}& \\ \hline
\multicolumn{1}{|l|}{\textit{Semi-aggressive}}&{3.1/3.2/4/5/6.1/6.2/7.1/7.2/7.3}&{7.1/7.3}&{3.1/5}&&{6.2} \\ \hline
\multicolumn{1}{|l|}{\textit{Active}}&&&{7.1}&&{6.2/7.2} \\ \hline
\multicolumn{6}{|l|}{\textbf{Means of Attack}} \\ \hline 
\multicolumn{1}{|l|}{\textit{Server}}&{6.1/7.2}&{7.2}&{6.2}&& \\ \hline
\multicolumn{1}{|l|}{\textit{Network}}&{3.1/3.2/5/7.2/7.3}&{5/7.3}&{3.1/3.2/6.1/7.1}&&{6.2} \\ \hline
\multicolumn{1}{|l|}{\textit{Client app}}&{5}&{}&{5}&{5/6.1}& \\ \hline
\multicolumn{1}{|l|}{\textit{User}}&{3.1/4/7.2/7.3}&{7.2}&{4/7.2}&&{7.2} \\ \hline
\multicolumn{1}{|l|}{\textit{Client machine}}&{6.1/6.2/7.1}&{6.1/7.1}&{6.2/7.1}&& \\ \hline
\multicolumn{6}{|l|}{\textbf{Purpose of Attack} \cite{dodsecurity:ciatriad, huang2011adversarial}} \\ \hline 
\multicolumn{1}{|l|}{\textit{Confidentiality}}&{4/5/6.1/7.2}&&{4/6.2/7.2}&& \\ \hline
\multicolumn{1}{|l|}{\textit{Availability}}&{3.1/6.2/7.3}&{7.2}&{3.1}&&{6.2/7.2} \\ \hline
\multicolumn{1}{|l|}{\textit{Integrity}}&{3.2/5/6.1/6.2/7.1/7.2/7.3}&{5/6.1/7.1/7.2/7.3}&{3.2/5/6.1/7.1/7.2}&{5/6.1}& \\ \hline
\end{tabular}
\vspace{0.1in}
\caption{Overview matrix of sections covering ML paradigms and security domains; details in section 2.2}
\end{table*}

We agree with assessment of top conferences in \cite{rossow2012prudent:top6confs}\footnote{We use the same conference-ranking websites to first decide our list of top-level conferences:(1)Microsoft Academic Search - Top Conferences in Security \& Privacy\url{http://academic.research.microsoft.com/RankList?entitytype=3&topDomainID=2&subDomainID=2}, (2)Guofei Gu - Computer Security Conference Ranking and Statistic\url{http://faculty.cs.tamu.edu/guofei/sec_conf_stat.htm}, and (3) Jianying Zhou - Top Crypto and Security Conferences Ranking \url{http://icsd.i2r.a-star.edu.sg/staff/jianying/conference-ranking.html}. All 3 rankings have the same top 6, and because Crypto and Eurocrypt do not have papers within our focus, we decided on these 4: ACM CCS, IEEE S\&P(hereafter ``SP''), NDSS, and USENIX Security(hereafter ``Sec'' or ``USENIX'')}. We systematically went through all proceedings between 2008 and early 2016 of the top 6 network- and computer-security conferences to collect relevant papers. Because of KDD's early and consistent publication record on ML applications in Security, and its status as a top-level venue for ML and DM applications, we also include KDD's 2008-2015 proceedings. To demonstrate the wide-ranging research attention drawn to ML applications in security, we also added chosen selections from the workshop AISec, International Conference on Machine Learning(ICML), Neural Information Processing Systems(NIPS), and Internet Measurement Conference(IMC) papers between 2008-2015, mostly in the ``Future Development'' section. 

\subsection{ML System Designs in Security}
Figure 1 shows the generalization of ML system designs when applied in security, that emerged from our survey of the papers(the legend is on the figure's bottom left). In different use cases, the system components may embody different names, but their functionalities and positions are captured in the figure. For example: 
\begin{enumerate}[nosep]
\item \textbf{\emph{Knowledge base}} is baseline of known normality and/or abnormality, depending on use cases, they include but are not limited to blacklist(BL), whitelist(WL), watchlist; known malware signatures, system traces, and their families; initial set of malicious web pages; existing security policies or rules, etc..  
\item \textbf{\emph{Data sources}} are where relevant data is collected. They can be either off-line or live online data feed, e.g. malware traces collected after execution(off-line), URL stream(online).
\item \textbf{\emph{Training data}} are labeled data which are fed to classifiers in training. They can be standard research datasets, new data(mostly from industry) labeled by human, synthetic datasets, or a mix. 
\item \textbf{\emph{Pre-processor and feature extractor}} construct features from data sources, for example: URL aggregators, graph representations, SMTP header extractions, n-gram model builders.  
\end{enumerate}
Dynamic analyzer and static analyzer are used most often in malware-related ML tasks, and human feedback loop is added when the system's design intends to be semi-supervised or human-in-the-loop(HITL).   

\begin{table*}
\centering
\begin{tabular}{|l|l|l|l|l|l|} \hline
\multicolumn{1}{|l|}{}& {Supervised}& {Semi-supervised}& {Unsupervised}& {HITL}&{Game Theory} \\ \hline 
\multicolumn{6}{|l|}{\textbf{Attacker Type}} \\ \hline 
\multicolumn{1}{|l|}{\textit{Passive}}&{58(49\%)}&{7(5.9\%)}&{24(20\%)}&{2(1.7\%)}&{0(0\%)} \\ \hline
\multicolumn{1}{|l|}{\textit{Semi-aggressive}}&{18(15\%)}&{4(3.4\%)}&{3(2.5\%)}&{0(0\%)}&{1(0.85\%)} \\ \hline
\multicolumn{1}{|l|}{\textit{Active}}&{0(0\%)}&{0(0\%)}&{0(0\%)}&{0(0\%)}&{2(1.7\%)} \\ \hline
\multicolumn{6}{|l|}{\textbf{Means of Attack}} \\ \hline 
\multicolumn{1}{|l|}{\textit{Server}}&{4(3.4\%)}&{1(0.85\%)}&{1(0.85\%)}&{0(0\%)}&{0(0\%)} \\ \hline
\multicolumn{1}{|l|}{\textit{Network}}&{17(14.4\%)}&{4(3.4\%)}&{11(9.3\%)}&{0(0\%)}&{1(0.85\%)} \\ \hline
\multicolumn{1}{|l|}{\textit{Client app}}&{4(3.4\%)}&{0(0\%)}&{1(0.85\%)}&{2(1.7\%)}&{0(0\%)} \\ \hline
\multicolumn{1}{|l|}{\textit{User}}&{31(26\%)}&{2(1.7\%)}&{9(7.6\%)}&{0(0\%)}&{2(1.7\%)} \\ \hline
\multicolumn{1}{|l|}{\textit{Client machine}}&{20(17\%)}&{4(3.4\%)}&{5(4.2\%)}&{0(0\%)}&{0(0\%)} \\ \hline
\multicolumn{6}{|l|}{\textbf{Purpose of Attack} \cite{dodsecurity:ciatriad, huang2011adversarial}} \\ \hline   
\multicolumn{1}{|l|}{\textit{Confidentiality}}&{16(13.6\%)}&{0(0\%)}&{7(6\%)}&{0(0\%)}&{0(0\%)} \\ \hline
\multicolumn{1}{|l|}{\textit{Availability}}&{9(7.6\%)}&{1(0.85\%)}&{5(4.2\%)}&{0(0\%)}&{3(2.5\%)} \\ \hline
\multicolumn{1}{|l|}{\textit{Integrity}}&{51(43.2\%)}&{10(8.5\%)}&{15(12.7\%)}&{2(1.7\%)}&{0(0\%)} \\ \hline
\end{tabular}
\vspace{0.1in}
\caption{Overview matrix of sections covering ML paradigms and security domains. Number in each box is the number of papers in our survey that belongs to the category, followed by a parenthesis with the percentage of the number among all counted papers.}
\end{table*}

\subsection{ML Paradigms in Security Problems}
Table 1 shows a matrix with rows indicating different ways of classifying the security problems, and the columns showing well-understood ML paradigms. Based on the threat models and modeling purposes presented in the papers, we qualitatively group the attacker into three groups. If there are multiple attacker types in one section, the section's numbering appears multiple times accordingly.
\begin{enumerate}[nosep]
\item \textbf{\emph{Passive}} attackers make no attempt to evade detections; their behaviors fit into descriptions of the threat models. 
\item \textbf{\emph{Semi-aggressive}} attackers have knowledge of the detectors, and only attempt to evade detections. 
\item \textbf{\emph{Active}} attackers do not only have knowledge of the detectors and attempt to evade detections, but also actively try to poison, mislead, or thwart detection.
\item \textbf{\emph{Knowledge}} of attackers, is the information in at least one of the five aspects: the learning algorithms themselves, the algorithms' feature spaces, the algorithm's parameters, training and evaluation data - regardless of being labeled or not - used by the algorithms, and decision feedback given by the algorithms \cite{huang2011adversarial, rndic2014practical, freeman2016prob:userauthen}.
\end{enumerate}
Influenced by \cite{dodsecurity:ciatriad,huang2011adversarial}, we extend their definitions, and qualitatively categorize attackers' primary purpose as to compromise \emph{confidentiality, availability} or \emph{integrity} of legitimate systems, services, and users. 
\begin{enumerate}[nosep]
\item Attacks on \textbf{\emph{confidentiality}} compromise the confidential or secret information of systems, services, or users (e.g. password crackers).
\item Attacks on \textbf{\emph{availability}} make systems and services unusable with unwanted information, requests, or many errors in defense schemes (e.g. DDoS, spam). 
\item Attacks on \textbf{\emph{integrity}} masquerade maliciously intentions as benign intentions in systems, services, and users (e.g. malware).
\end{enumerate}
We also define ML paradigms shown in the matrix:
\begin{enumerate}[nosep]
\item \textbf{\emph{Supervised}} learning uses labeled data for training. 
\item \textbf{\emph{Semi-supervised}} learning uses both labeled and unlabeled data for training. 
\item \textbf{\emph{Unsupervised}} learning has no labeled data available for training.
\item \textbf{\emph{Human-in-the-loop(HITL)}} learning incorporates active human feedback to algorithm's decisions into the knowledge base and/or algorithms.
\item \textbf{\emph{Game Theory(GT)}}-based learning considers learning as a series of strategic interactions between the model learner and actors with conflicting goals. The actors can be data generators, feature generators, chaotic human actors, or a combination\cite{grosshans2013bayesian, rndic2014practical, yan2012towards, huang2011adversarial, Laskov:2009:quantitaviesecanalysis:ML}.
\end{enumerate}

For "Means of Attacks" in Table 1, server, network, and user are straightforward and intuitive, so here we only describe ``client app'' and ``client machine''. \textbf{\emph{Client app}} is any browser-based means of attack on any client device, and \textbf{\emph{client machine}} is any non-browser-based means of attack on any client device.

As shown in Table 1, the majority of surveyed papers in different security domains use supervised learning to deal with passive or semi-aggressive attackers. However, the core requirement of supervised learning - labeled data - is not always viable or easy to obtain, and authors have repeatedly written about the difficulty of obtaining labeled data for training. Based on this observation, we conclude that \emph{exploring semi-supervised and unsupervised learning approaches would expand the research foundation of ML applications in security domains, because semi-supervised and unsupervised learning can utilize unlabeled datasets which had not been used by supervised learning approaches before.}

Moreover, during our survey, we realized that many ML applications in security assume that training and testing data come from the same distribution (in statistical terms, this is the assumption of stationarity). However, in the real world, it is highly unlikely that data are stationary, let alone that the data could very well be generated by an adversarial data generator producing training and/or testing data sets, as the case in \cite{grosshans2013bayesian}, or simply be generated responding to specific models as in \cite{Bruckner:2011kdd:spamfilter}. Our observation from the comprehensive survey confirmed \cite{huang2011adversarial}'s statement, and we propose that \emph{GT-based learning approaches and HITL learning system designs should be explored more, in order to design more efficient security defense mechanisms to deal with active and unpredictable adversaries. At the same time, human knowledge and judgment in HTIL should go beyond feature engineering, to providing feedback to decisions made by ML models}. Some theory-leaning papers have modeled spam filtering as Bayesian games or Stackelberg games\cite{grosshans2013bayesian, Bruckner:2011kdd:spamfilter}. Use cases in data sampling, model training with poisoned or low-confidence data have also been briefly explored in literature\cite{biggio2013data, yan2012towards, biggio2014poisoning}.

\subsection{Timeline of ML in Security} 
Based on seminal works and establishments in notable venues, the gradually increasing levels of interest in ML research applied to Security is fairly visible. Here we gathered some milestone events:
\begin{enumerate}[nosep]
\item 1987: Denning published ``An Intrusion Detection System'' \cite{Denning:1987ids}, first framing security as a learning problem
\item 1998: DARPA IDS design challenge\cite{DARPAIDS1998}
\item 1999: KDD Cup IDS design challenge\cite{kdd1999:data, kdd1999:result}
\item 2008: CCS hosted the 1st AISec workshop. Continues to operate each year\cite{CCS:AISec2016TOpicScope}
\item 2007, 2008: Twice, KDD hosted the International Workshop on Privacy, Security, and Trust(PinKDD)\cite{pinkdd2008:reports}
\item 2010, 2012: Twice, KDD hosted Intelligence and Security Informatics Workshop(ISI)\cite{kdd2010isi:workshop, kdd2012isi:workshop} 
\item 2011: ``Adversarial Machine Learning'' published in 4th AISec\cite{huang2011adversarial}
\item 2012: ``Privacy and Cybersecurity: The Next 100 Years'' by Landwehr et al published\cite{landwehr2012privacy}
\item 2013: Manifesto from Dagstuhl Perspectives Workshop published as ``Machine Learning Methods for Computer Security'' by Joseph et al \cite{dagstuhlmanifesto:2013}.
\item 2014: KDD hosted its 1st ``Security \& Privacy'' session in the main conference program\cite{kdd2014:spsession}
\item 2014: ICML hosted its 1st, and so far the only workshop on Learning, Security, and Privacy(LSP)\cite{ICML2016:securityworkshop}
\item 2016: AAAI hosted its 1st Artificial Intelligence for Cyber Security workshop(AISC)\cite{AAAI:AISC2016}
\end{enumerate}

\subsection{Related Work}
Despite the surge of research interests and industry applications in the intersection of ML and security, few surveys or overviews were published after 2008, the watershed year of increasing interest in this particular domain. In 2013 \cite{relatedwork2013:serverside:webapp}surveyed server-side web application security, \cite{aggarwal2013application} surveyed data mining applied to security in the cloud focusing on intrusion detection, \cite{bhattacharyya2013network} discussed an ML perspective in network anomaly detection. While they are helpful and informative, the former two are limited by their scope and perspective, and the latter serves as a textbook, hence absent the quintessential of survey - mapping the progresses and charting the state-of-art. A collection of papers in 2002 and 2012 \cite{jajodin20022012:DMinCompSec} discussed applications of DM in computer security, but lacks a systematic survey on ML applications in resolving security issues. \cite{ahmed2007machine} briefly compared two network anomaly detection techniques, but limited in scope. \cite{Chandola:2009:ADS} of 2009 conducted a comprehensive survey in anomaly detection techniques, some involving discussions of security domains. The Dagstuhl Manifesto in 2013 \cite{dagstuhlmanifesto:2013} articulated the status quo and looked to the future of ML in security, but the majority of the literature listed were published before 2008. \cite{sommer2010:paxson} of 2010 highlighted use cases and challenges for ML in network intrusion detection, but did not incorporate a high-level review of ML in security in recent years.  

\section{NETWORK SECURITY}
\subsection{Botnets and Honeypots}
Research works on botnets among our surveyed literature focuses mainly on designing systems to detect command-and-control(C\&C) botnets, where many bot-infected machines are controlled and coordinated by few entities to carry out malicious activities\cite{gu2008botsniffer:ndss, Jacob2011:jackstraw:bottraffic, Nagaraja:2010BotGrep}. Those systems need to learn decision boundaries between human and bot activities, therefore ML-based classifiers are at the core of those systems, and are often trained by labeled data in supervised learning environments. The most popular classifier is support vector machines(SVMs) with different kernels, while spatial-temporal time series analysis and probabilistic inferences are also notable techniques employed in ML-based classifiers. Topic clustering, mostly seen in natural language processing(NLP), is used to build a large-scale system to identify bot queries \cite{zhang2013intention:ndssbotnet}. In botnet detection literature, 3 core assumptions are widely shared: 
\begin{enumerate}[nosep]
\item Botnet protocols are mostly C\&C \cite{gu2008botsniffer:ndss, usenix2012:nbsvm:classifycrawler, Jacob2011:jackstraw:bottraffic}
\item Individual bots within same botnets behave similarly and can be correlated to each other \cite{gu2008botminer, usenix2012:nbsvm:classifycrawler}
\item Botnet behaviors are different and distinguishable from legitimate human user, e.g. human behaviors are more complex\cite{gianvecchio2008measurement, hu2009rbbotnets:ndss,lee2016you:measurement:bot}
\end{enumerate}
Other stronger assumptions include that bots and humans interact with different server groups \cite{chen2011detectingbots:kdd}, and content features from messages generated by bots and human are independent \cite{gianvecchio2008measurement}. While classification techniques differ, WLs, BLs, hypothesis testing, and a classifier \cite{gu2008botsniffer:ndss, gianvecchio2008measurement, Nagaraja:2010BotGrep} are usual system components. Attempts have been made to abstract state machine models of network to simulate real-world network traffic and create honeypots \cite{krueger2012learning:ccs}. Ground truths are often heuristic \cite{gu2008botminer}, labeled by human experts, or a combination - even at large scale, human labeled ground truths are used, for example in \cite{lee2016you:measurement:bot}, game masters' visual inspections serve as ground truth to detect bots in online games. In retrospect, the evolution of botnet detection is clear: from earlier and more straightforward uses of classification techniques such as clustering and NB, the research focus has expanded from the last step of classification, to the important preceding step of constructing suitable metrics, that measures and distinguishes bot-based and human-based activities\cite{zhang2013intention:ndssbotnet, lee2016you:measurement:bot}.  

\subsection{Proxies and DNS}
Classifying DNS domains that distribute or host malware, scams, and malicious content has drawn research interest especially in passive DNS analysis. There are two main approaches: reputation system\cite{antonakakis2010building:dynamicRepsys:DNS, antonakakis2011detecting, qiu2009locating:prefixhijacking} and classifier\cite{bilge2011exposure:j48dt, song2009:maliciousWebquery}. Reputation system scores benign and malicious domains and DNS hosts, and a ML-based classifier learns boundaries between the two. Nonetheless, both reputation system and classifier use various decision trees, random forest(RF), na\"{\i}ve Bayes(NB), SVM, and clustering techniques for mostly supervised learning-based scoring and classification. Many features used are from protocols and network infrastructures, e.g. border gateway protocol(BGP) and updates, automated systems(AS), registration, zone, hosts, and public BLs. Similar to botnet detectors, variations of BL, WL, and honeypots\cite{small2008:nlpelicit:maliciouspayload} are used in similar functions as knowledge bases, while ground truths are often taken from public BLs, limited WLs, and anti-virus(AV) vendors such as McAfee and Norton \cite{antonakakis2011detecting, bilge2011exposure:j48dt, antonakakis2010building:dynamicRepsys:DNS}. But before any ML attempts take place, most studies would assume the following:
\begin{enumerate}[nosep]
\item Malicious uses of DNS are distinct and distinguishable from legitimate DNS services.
\item The data collection process - regardless of different names such as data flow, traffic recorder, or packet assembler - follows a centralized model. In other words, all the traffic/data/packets flow through certain central node or nodes to be collected. 
\end{enumerate}
Stronger assumptions include that AS hijackers cannot manipulate AS path before it reaches them\cite{qiu2009locating:prefixhijacking}, and maliciousness will trigger an accurate IP address classifier to fail\cite{venkataraman2013automatically}. Besides analyzing the status quo, \cite{fatemieh2011using:measurement:networkhost, venkataraman2013automatically, vissers2015parking:domainsRF} showed efforts to preemptively protect network measurement integrity and predict potentially malicious activities from web domains and IP address spaces. 

\section{Security Services}
Both offense and defense for access control, authentication, and authorization reside within the domain of Security Services. Defeating audio and visual CAPTCHAs(Completely Automated Public Turing test to tell Computers and Humans Apart)\cite{golle2008machine:asirracaptcha, Bursztein2011:textbasedCAPTCHA:strongandweak, Bursztein2011:textbasedCAPTCHA:strongandweak,Gao:2013:attackHollowCaptcha, gao2016simple}, cracking passwords\cite{weir2009password, chatterjee2015cracking, ma2014study}, measuring password strengths\cite{castelluccia2012adaptive,chatterjee2015cracking, kelley2012guess:passwordstrengh}, and uncovering anonymity\cite{zander2008improved:samplinghiddenservice,Afroz:stylometry_lr,narayanan2012feasibility:authorship,dyer2012peek:sp:taattack} are 4 major use cases. On the offense, specialized ML domains such as computer vision, signal processing, and NLP automate attacks on user authentication services i.e. textual or visual passwords and CAPTCHAs, and uncover hidden identities and services. On the defense side, entropy-based and ML-based systems calculate password strengths. Other than traditional user authentication schemes, behavioral metrics of users are also introduced. Following the generalized ML pipeline shown in Figure 1, the ``classifier'' is replaced by ``recognition engine'' in the password cracking process, and ``user differentiation engine'' in authentic metric engineering \cite{Zheng2011:svmuser:mouse, frank2008prob:model:rbac}. Hence the process becomes: ``Data source $\rightarrow$ Pre-process \& feature extraction $\rightarrow$ Recognition or user differentiation engine $\rightarrow$ Decision'' for ML-based security services. A noteworthy trend to observe, is that attacks on CAPTCHAs are getting more generalized - from utilizing SVM in 2008 to attack a specific type of text CAPTCHA\cite{golle2008machine:asirracaptcha}, in 2015 a generic attach approach to attack text-based CAPTCHA \cite{gao2016simple} was proposed.  

ML-based attacks on textual and visual CAPTCHA typically follow the 4-step process:  
\begin{enumerate}[nosep]
\item \emph{Segmentation}: e.g. signal to noise ratio(SNR) for audio; hue, color, value(HSV) for visual \cite{golle2008machine:asirracaptcha, bursztein2011failure, Gao:2013:attackHollowCaptcha, gao2016simple}
\item \emph{Signal or image representation}: e.g. discrete Fourier transformation(audio)\cite{bursztein2011failure}, letter binarization(visual) \cite{gao2016simple}
\item \emph{Feature extraction}: e.g. spectro-temporal features, character strokes \cite{golle2008machine:asirracaptcha, bursztein2011failure}
\item \emph{Recognition}: K-nearest neighbor(KNN), SVM(RBF kernel), convolutional neural networks(CNN) \cite{Bursztein2011:textbasedCAPTCHA:strongandweak, weir2009password}
\end{enumerate}

On the side of password-related topics in security services, there are 2 password models: whole-string Markov models, and template-based models \cite{ma2014study}. Concepts in statistical language modeling, such as natural language encoder and n-grams associated with Markov models(presented as directed graphs with nodes labeled by n-grams), and context-free grammars are common probabilistic foundations to build password strength meters and password crackers \cite{weir2009password, bursztein2011failure, chatterjee2015cracking}. 

\section{SOFTWARE \& APPLICATION SECURITY}
ML research in software and applications security mostly concentrate on web application security in our survey, and have used supervised learning to train popular classifiers such as NB and SVM to detect web-based malware and JavaScript(JS) code\cite{Curtsinger:2011:malware, Kapravelos13:malware:KNN}, filter unwanted resources and requests such as malicious advertisements\cite{sculley2011detectingad:adml, Bhagavatula2014:mlresourcefiltering, zhang2008highly, lu2011surf}, predict unwanted resources and requests(e.g. future blacklisted websites)\cite{usenix2014badpageml, usenix2015:imageobjrecog:webdeface, robertson2010effective}, and quantify web application vulnerabilities\cite{Chapman2011:sidechannel:vulnerabilitydetector:webapp}. While \cite{robertson2010effective} explored building web application anomaly detector with scarce training data, most use cases follow the supervised paradigm assuming plentiful labeled data: Data source(web applications, static/dynamic analyzers) $\rightarrow$ feature extraction(often with specific pre-filter, metrics, and de-obfuscator if needed) $\rightarrow$ classifiers trained with labeled data. Apart from this supervised setting, if a human expert's feedback is added after classifiers' decisions\cite{sculley2011detectingad:adml}, it forms a semi-supervised system. Regardless of system designs, the usual assumption holds: \emph{malicious activities or actors are different from normal and benign ones likely do not change much}. The knowledge bases of normality and abnormality can vary, from historical regular expression lists\cite{Bhagavatula2014:mlresourcefiltering} to other publicly available detectors\cite{Kapravelos13:malware:KNN}. Graph-based algorithms\cite{zhang2008highly} and image recognition\cite{usenix2015:imageobjrecog:webdeface} are both used in resource filtering, but in detecting JS malware and evasions and quantifying leaks, having suitable measurements of similarities is a significant focal point. Indeed, from \cite{Curtsinger:2011:malware, Kapravelos13:malware:KNN, Chapman2011:sidechannel:vulnerabilitydetector:webapp}, ML-based classifiers do well in finding similarities between mutated malicious code snippets, while the same code pieces could evade static or dynamic analyzer detections. 

\section{SYSTEM SECURITY}
\subsection{Vulnerability and Policy Management}
As Landwehr noticed\cite{Landwehr:2008keynote}, ML can be applied in SPM. However, in automatic fingerprinting of operating systems(OS), C4.5 decision tree, SVM, RF, KNN - some most commonly used ML-based classifiers in security - failed to distinguish remote machine instances with coarse- and fine-grained differences, as the algorithms cannot exploit semantic knowledge of protocols or send multi-packet probes \cite{richardson2010limits}. Yet by taking advantage of semantic and syntactic features, plus semi-supervised system design, \cite{rasthofer2014machine, WHYPER:usenix2013:nlp:spm, usenix2015:semisup:EASEAndroid} showed that SVM(optimized by sequential minimal optimization[SMO] algorithm), KNN, and NLP techniques do well in Android SPM. On the other hand, in vulnerability management, \cite{makanju2009clustering, bozorgi2010beyond, peng2012using:risksofandroid, yamaguchi2015automatic, usenix2015:rfforecase:incident}, clustering techniques have done well in predicting future incidents and infer vulnerability patterns in code, as well as NB, SVM, and RF in ranking risks and identifying proper permission levels. Both vulnerability management and SPM also focus on devising proper metrics for ML applications: from heuristics based on training set \cite{peng2012using:risksofandroid}, Jaro distance \cite{yamaguchi2015automatic}, to outside reputation system oracles \cite{bozorgi2010beyond}, metrics are needed to compare dependency graphs, string similarities, and inferred vulnerability patterns. In most use cases, because of the need for labeled data to train supervised learning systems, many systems follow the generalized training process in Figure 1: ``Knowledge base $\rightarrow$ offline trainer $\rightarrow$ online or offline classifier''. When policy management decisions need feedback, a HITL design is in place where end human users' feedback is directed to knowledge base. One distinguishing tradition in ML applications research in this domain, is a strong emphasis on measurement - selecting or engineering proper similarity or scoring metrics are often important points of discussion in research literature. From earlier uses of heuristics in clustering algorithms, to more recent semantic connectivity measurement applied in semi-supervised systems, both the metrics and the system designs for vulnerability and security policy management have evolved to not only identify, but also to infer and predict future vulnerable instances. 

\subsection{Information Flow and DDoS}
Compared to other security domains, ML research in information flow and DDoS focus more on evasion tactics and limits of ML systems in adversarial environments. Hence we grouped together the two sub-domains, and marked studies in Table 7 with ``(IF)'' and ``(DDoS)'' accordingly. For DDoS\cite{berral2008adaptive:againstddos, venkataraman2008limits, yan2012towards}, the usual assumption is that \emph{patterns of attack and abuse traffic are different from normal traffic }\cite{berral2008adaptive:againstddos}, but \cite{venkataraman2008limits} challenged it by proposing an adversary who can generate attributes that look as plausible as actual attributes in benign patterns, and caused failure in ML-based automated signature generation to distinguish benign and malicious byte sequences. Then, \cite{yan2012towards} introduced GT to evaluate DDoS attack and defense in real-world. For information flow\cite{vsrndic2013detection, rndic2014practical,lu2015checking,xu2016automatically}, assumptions can take various forms. In PDF classifiers based on document structural features, it is \emph{malicious PDF has different document structures than good PDFs} \cite{vsrndic2013detection}; in Android privacy leak detector, it is \emph{the majority of an Android application's semantically similar peers has similar privacy disclosure scenarios}\cite{lu2015checking}. But \cite{rndic2014practical} poses semi-aggressive and active attackers with some information about the data, feature sets, and/or algorithms, and then attackers successfully evade ML-based PDF classifiers. Another example is, PDF malware could be classified \cite{vsrndic2013detection}, and then a generic and automated evasion technique based on genetic programming is successfully experimented\cite{xu2016automatically}. Overall, while using SVM, RF, and decision trees trained with labeled data to detect and predict DDoS and malicious information and data flows, ML applications in information flow and DDoS challenge the usual assumption of stationary adversary behaviors. From collecting local information only, to proposing a general game theory-based framework to evaluate DDoS attacks and defense, and from using static method to detect malicious PDF file to generic automated evasion, the scope of ML applications in both DDoS and IF have expanded and generalized over the years. 

\section{MALWARE, SOCIAL ENGINEERING \& IDS}
\subsection{Malware Detection and Mitigations}
Program-centric or system-centric, there are 3 areas that draw most ML application research attention in malware: malware detection\cite{tamersoy2014guilt, invernizzi2014nazca, arp2014drebin, usenix2015:lmt:malwaredev, usenix2015:rnn:binary:function, smutz2016tree, usenix2012:clustermalware, usenix2012:svm:osn:socware, Lanzi2010:accessminer:malware:ngram, ye2011combining, Kolbitsch2009:malware:dag:graphmatching,ahmed2009using:spatialtemporal:malware}, classifying unknown malware into families\cite{kong2013discriminant:metriclearning, borgolte2013delta, biggio2014poisoning, Jang2011:featurehashing:malwaresemantics:bitshred, ye2010automatic, bayer2009scalable}, and auto-extract program or protocol specifications \cite{Fredrikson:2010malwaresmanl,Comparetti2009:prospex, kirat2015malgene:idf:longestsubsequence}. Realizing the signature and heuristic-based malware detectors can be evaded by obfuscation and polymorphism \cite{Kolbitsch2009:malware:dag:graphmatching}, more behavior-based matching and clustering systems and algorithms have been researched. Figure 1 already shows a generalized ML system design for malware detection and classification, and a more detailed description is below: 
\begin{enumerate}[nosep]
\item Collect malware artifacts and samples, analyze them, execute them in a controlled virtual environment to collect traces, system calls, API calls, etc. \cite{Lanzi2010:accessminer:malware:ngram, ahmed2009using:spatialtemporal:malware, Jang2011:featurehashing:malwaresemantics:bitshred} \footnote{We direct our readers for more details in \cite{rossow2012prudent:top6confs}, which evaluated rigor and prudence of academic research up to 2011 that rely on malware execution, and provided a guide rubric for safety for handling malware datasets}. Or, directly use information from already completed static and/or dynamic analyses.
\item Decide or devise similarity measurements between generalized binaries, system call graphs(SCG), function call graphs(FCG), etc., then extract features \cite{bayer2009scalable, Comparetti2009:prospex, borgolte2013delta, kong2013discriminant:metriclearning}
\item Classify malware artifacts into families in-sample, or cluster them with known malware families. The classifiers and clustering engines are usually trained with labeled data\cite{Kolbitsch2009:malware:dag:graphmatching, ye2010automatic, antonakakis2010building:dynamicRepsys:DNS}. Popular ones are SVM and RF for classification, and hidden Markov model(HMM) and KNN alongside different clustering techniques. 
\end{enumerate}
Even in the use case of auto-extract specifications, supervised learning with labeled data is needed when behavior profiles, state machine inferences, fuzzing, and message clustering are present. Evasion techniques of detectors and poisoning of ML algorithms are also discussed, and typical evasion techniques include obfuscation, polymorphism, mimicry, and reflecting set generation\cite{smutz2016tree, rndic2014practical}. Malware detection and matching based on structural information and behavior profiles\cite{Kolbitsch2009:malware:dag:graphmatching,kong2013discriminant:metriclearning, bayer2009scalable} show a tendency to use graph-based clustering and detection algorithms, and similarity measurement used in these algorithms have ranged from Jaccard distance to new graph-based matching metrics. While clustering techniques have been mostly used in malware detection, a nearest neighbor technique is explored to evade malware detection. 

\subsection{Social Engineering: Phishing, Malicious Content and Behaviors}
Spams, malicious webpages and URLs that redirect or mislead un-suspecting users to malware, scams, or adult content \cite{vissers2015parking:domainsRF} is perhaps as old as civilian use of the Internet. Research literature mostly focus on 3 major areas: detecting phishing malicious URLs\cite{ma2009identifyingURL, ma2009beyondURL:kdd, Blum2010:phishingurl:lexicalfeatures, whittaker2010large, lee2012warningbird, zhao2013cost:malurl:2013kdd}, filtering spam or fraudulent content \cite{chatterjee2008robust:repsys, hao2009detecting:reputationsys, thomas2011design, Bruckner:2011kdd:spamfilter, afroz2012detecting:writingstyle, grosshans2013bayesian, invernizzi2012evilseed, usenix2013:bayesian:spam:phonenumber, zhang2014dspin:measurement:nlp, whalen2014model}, and detecting malicious user account behaviors\cite{egele2013compa:ndssnlp, usenix2014:anomuserbehavior, boshmaf2015integro, usenix2015:ml:malicioususer, usenix2015:adml:crowdsourcer:svmrf}. Moreover, because phishing \footnote{We agree with \cite{whittaker2010large}'s definition of phishing: ``without permission, alleges to act on behalf of a third party with the intention of confusing viewers into performing an action with which the viewer would only trust a true agent of the third party''} is a classic social engineering tactic, it is often the gateway of many studies to detect malicious URLs, spam, and fraudulent content. To identify malicious URLs, ML-based classifiers draw features from webpage content(lexical, visual, etc.), URL lexical features, redirect paths, host-based features, or some combinations of them. Such classifiers usually act in conjunction with knowledge bases which are usually in-browser URL BLs or from web service providers. If the classifier is fed with URL-based features, it is common to set an URL aggregator as a pre-processor before extracting features. Mostly using supervised learning paradigm, NB, SVM with different kernels, and LR are popular ML classifiers for filtering spam and phishing. Meanwhile, GT-based learning to deal with active attackers is also evaluated in spam filtering. \cite{grosshans2013bayesian} evaluates a Bayesian game model where the defense is not fully informed of the attacker's objectives and the active adversary can exercise control over data generation, \cite{Bruckner:2011kdd:spamfilter} proposes a Stackelberg game where spammer reacts to the learner's moves. Stronger assumptions also exist: for example, \cite{usenix2013:bayesian:spam:phonenumber} assumes spammers' phone blocks follow a beta distribution as conjugate prior for Bernoulli and binomial distribution. Another social engineering tactic is spoofing identities with fake or compromised user accounts, and detection of such malicious behaviors utilize features from user profiles, spatial-, temporal-, and spatial-temporal patterns, and user profiles are used in particular to construct normality. Graph representation and trust propagation models are also deployed to distinguish genuine and malicious accounts with different behavior and representations\cite{boshmaf2015integro, usenix2015:ml:malicioususer, usenix2015:adml:crowdsourcer:svmrf}. Tracing the chronology of applying ML to defend against social engineering, one trend is clear: while content-, lexical-, and syntactic-based features are still being widely used, constructing graph representations and exploring temporal patterns of redirect paths, events, accounts, and behaviors have been on the rise as feature spaces for ML applications in defend against social engineering efforts. Accordingly, the ML techniques have also changed from different classification schemes to graphic models. It is also noteworthy that in \cite{Bruckner:2011kdd:spamfilter, usenix2015:adml:crowdsourcer:svmrf}, addressing adversarial environments' challenges to ML systems is elaborated as primary research areas, instead of a short discussion point.

\subsection{IDS}
From feature sets to algorithms and systems, IDS has been extensively studied. However, as \cite{symons2012nonparametric} cautioned, ML can be easily conflated with anomaly detection. While both are applied to build IDS, important difference is that ML aims to generalize expert-defined distinctions, but anomaly detection focuses on finding unusual patterns, while attacks are not necessarily anomalous. For example, \cite{wressnegger2013close} distinguished n-gram model's different use cases: anomaly detection uses it to construct normality(hence more appropriate when no attack is available for learning), and ML classifiers learn to discriminate between benign and malicious n-grams(hence more appropriate when more labeled data is present). Since 2008, works at top venues have added to the rigor for ML applications in IDS. For example, a common assumption of IDS is: \emph{Anomalous or malicious behaviors or traffic flows are fundamentally different from normal ones}, but \cite{symons2012nonparametric} challenges the assumption by studying low-cardinality intrusions where attackers don't send a large number of probes. To address adversarial learning environment and minimal labels in training data, semi-supervised paradigms, especially active learning, are also used\cite{symons2012nonparametric, Gornitz:2009:activeLearningIDS}. Heterogeneous designs of IDS in different use cases give rise to many ad-hoc evaluations in research works, and a reproducibility and comparison framework was proposed to address the issue\cite{juba2015principled}. Meanwhile, techniques such as graph-based community detection\cite{ding2012intrusion:measurement:ids}, time series-based methods\cite{xie2012review, momtazpour2015analyzing}, and generalized support vector data description in cyber-physical system and adversarial environment for auto-feature selection\cite{Kloft2008:autofeaturesselection}, have also emerged. Although they carry different assumptions of normality and feature representations, the supervised ML system design remains largely the same. Besides the fact the more techniques and use cases have been proposed, the focus of research in IDS had evolved from discovering new techniques and use cases, to rigorously evaluating fundamental assumptions and workflows of IDS. For example, while feature selection has stayed as a major component, there are re-examination of assumptions and measurements on what constitutes normality and abnormality\cite{ding2012intrusion:measurement:ids}, alternative to more easily acquire data and use low-confidence data for ML systems\cite{Gornitz:2009:activeLearningIDS}, and proposal on validating reproducibility of results from different settings\cite{juba2015principled}. 

\section{FUTURE DEVELOPMENT}
One key goal of our SoK survey is to help researchers look into the future. ML applications in security domains are attracting academic research attention as well as industrial interest, and this presents a valuable opportunity for researchers to navigate the landscapes between ML theories and security applications. There are also opportunities to explore if there are some types of ML paradigms that are especially well suited to particular security problems.  Apart from highlighting that 1) semi-supervised and unsupervised ML paradigms are more effective in utilizing unlabeled data, hence ease the difficulty of obtaining labeled data, and 2) GT-based ML paradigms and HITL ML system designs will become more influential in dealing with semi-aggressive and aggressive attackers, we also share the following seven speculations of future trends, based on our current SoK. 

\begin{enumerate}[nosep]
\item \textbf{Metric Learning}: Measurement has become more and more conspicuous for ML research in security, mostly in similarity measurement for clustering algorithms\cite{gianvecchio2008measurement, fatemieh2011using:measurement:networkhost,zander2008improved:samplinghiddenservice, kong2013discriminant:metriclearning}. Proper measurements and metrics are also used to construct ground truths to evaluate ML-based classifiers, and also have important roles in feature engineering\cite{zhang2014dspin:measurement:nlp, lee2016you:measurement:bot, holz2008measuring:metriclearning:network, rubinstein2009antidote}. Given the ubiquitous presence of metrics and the complex nature of constructing them, ML applications in security will benefit much from metric learning.

\item \textbf{NLP}: Malicious content, spam, and malware analysis and detections have used tools from statistical language modeling(e.g. n-gram-based representation for strings in code and HTTP request)\cite{song2009:maliciousWebquery, castelluccia2012adaptive, moskovitch2008acquisition, Lanzi2010:accessminer:malware:ngram, invernizzi2012evilseed, whalen2014model}, As textual information explodes, NLP will become more widely used beyond source filtering and clustering e.g. \cite{krueger2012learning:ccs} use n-gram models to infer state machines of protocols.  

\item \textbf{Upstream movement of ML in security defense designs}. In malware detection and classifications, behavior- and signature-based malware classifiers have used inputs from static and dynamic binary analysis as features\cite{Comparetti2009:prospex, ye2010automatic, arp2014drebin, usenix2015:lmt:malwaredev}, and \cite{usenix2015:rnn:binary:function} already shows RNN can be applied to automatically recognize functions in binary analysis. We also see ML algorithms applied in vulnerability, device, and security policy management, DDoS mitigation, information flow quantifications, and network infrastructure\cite{berral2008adaptive:againstddos, lu2015checking, whalen2014model, Lanzi2010:accessminer:malware:ngram}. Hence, it is reasonable to expect that more ML systems and algorithms will move upstream in more security domains. 

\item \textbf{Scalability}: With increasing amount of data from growing numbers of information channels and devices, scale of ML-based security defenses will become a more important aspect in researching ML applications in security \cite{invernizzi2014nazca, zhang2013intention:ndssbotnet, whittaker2010large, panchenko2016website}. As a result, large-scale systems will enable \textbf{distributed graph algorithms} in malware analysis, AS path hijacker tracing, cyber-physical system fault correlation, etc..\cite{Nagaraja:2010BotGrep, chen2011detectingbots:kdd, chatterjee2015cracking, WHYPER:usenix2013:nlp:spm, lu2015checking, Kolbitsch2009:malware:dag:graphmatching} 
\item \textbf{Specialized probabilistic models} will be applied beyond the context of classifiers, e.g. access control\cite{frank2008prob:model:rbac}.

\item High FP rates have always been a concern for system architects and algorithm researchers \cite{juba2015principled, Bhagavatula2014:mlresourcefiltering}. \textbf{Reducing FP rates} will grow from an ad-hoc component in various system designs, to independent formal frameworks, algorithms, and system designs.

\item \textbf{Privacy enforcement} was framed as a learning problem recently in \cite{usenix2014:privacyenforcement:learningproblem}, in the light of many publications on privacy-preservation in ML algorithms, and privacy enhancement by probabilistic models\cite{johnson2013privacy,wang2013rbridge, cao2015towards, huang2015genoguard, kumar2009mining, shokri2011quantifying}. This new trend will become more prominent.

\end{enumerate}

\section{Conclusion}
In this paper, we analyzed ML applications in security domains by surveying literature from top venues of our field between 2008 and early 2016. We attempted to bring clarity to a complex field with intersecting expertises by identifying common use cases, generalized system designs, common assumptions, metrics or features, and ML algorithms applied in different security domains. We constructed a matrix showing the intersections of ML paradigms and three different taxonomy structures to classify security domains, and show that while much research has been done, explorations in GT-based ML paradigms and HITL ML system designs are still much desired (and under-utilized) in the context of active attackers. We point out 7 promising areas of research based on our observations, and argue that while ML applications can be powerful in security domains, it is critical to match the ML system designs with the underlying constraints of the security applications appropriately. 

\section{Acknowledgment}
We would like to thank Megan Yahya, Krishnaprasad Vikram, and Scott Algatt for their time and valuable feedback. 

\appendix

\begin{table*} 
\footnotesize
\caption{Botnet Detectors} 
\footnotesize
\begin{tabular}{|p{2.8cm}|p{4cm}|p{5.1cm}|p{5.1cm}|} \hline
Study & Goal & System Components & ML Techniques \\ \hline
\texttt Gu NDSS'08\cite{gu2008botsniffer:ndss}& Detect hosts of botnets and malicious patterns& WL, feature extractor, clustering engine & X-Means clustering  \\ \hline

\texttt Gianvecchio NDSS'08\cite{gianvecchio2008measurement}& Distinguish between human- and bot-generated chats & Entropy-based classifier, ML-based classifier  & NB  \\ \hline 

\texttt Gu USENIX'08\cite{gu2008botminer}& Network-based anomaly detection for botnet C\&C channels & WL, watch list, protocol matcher, activity and message response detector, spatial-temporal correlation engine & sequential probability ration testing(SPRT), hierarchical clustering \\ \hline

\texttt Hu NDSS'09\cite{hu2009rbbotnets:ndss}& Detect redirection botnets&  BL, WL, resource filtering(content analysis, URL probing, network analysis, IP correlation), classifier & Linear SVM, SPRT \\ \hline

\texttt Nagaraja USENIX'10\cite{Nagaraja:2010BotGrep}& Find peer-to-peer botnets by traffic patterns& Traffic monitor honeypot communications, background traffic collector, graph preprocess, inference system & Graph clustering \\ \hline

\texttt Chen KDD'11\cite{chen2011detectingbots:kdd}&  Classify bots or human interactions& Flow parser, BL, graph feature extractor, online classifier & Least squares SVM \\ \hline

\texttt Jacob USENIX'11\cite{Jacob2011:jackstraw:bottraffic}& Finding C\&C connections from bot traffic & Host-based information, C\&C behavioral graphs, graphs of known C\&C connections & Graph clustering(metric: non-induced maximum common subgraph)  \\ \hline

\texttt Jacob USENIX'12\cite{usenix2012:nbsvm:classifycrawler}& Detect and contain malicious web crawlers& BL, WL, classifier& Ensemble classifier: NB, SVM(Gaussian radial basis function ``RBF'' kernel), association rules  \\ \hline

\texttt Krueger CCS'12\cite{krueger2012learning:ccs}& Infer a state machine and message format of a protocol from network traffic & Message embedder, cluster, inference engine, template engine & N-grams, tokens, binomial test(to reduce dimensionality), non-negative matrix factorization(NMF), Markov model \\ \hline

\texttt Zhang NDSS'13 \cite{zhang2013intention:ndssbotnet}& Study bot searches on a search engine& Feature extraction, pattern tree generation, topic clustering, classifier & Spectral clustering(metric: Jaccard distance), single linkage hierarchical clustering(SLHC) \\ \hline 

\texttt Lee NDSS'16 \cite{lee2016you:measurement:bot}& Detect game bots in massively multiplayer online role-playing games& Game logs, transformation, feature extraction, metric computing, classifier & Logistic regression(LR), cosine similarity, Hurst exponent, exponential weighted moving average \\ \hline 

\end{tabular}
\end{table*}

\begin{table*}
\footnotesize
\caption{Proxies and DNS}
\begin{tabular}{|p{2.4cm}|p{4.9cm}|p{5.2cm}|p{4.5cm}|} \hline
Study & Goal & System Components & ML Techniques \\ \hline
\texttt Small USENIX'08\cite{small2008:nlpelicit:maliciouspayload}& Auto-generate responses to malicious requests and harness payloads& Data collection, pre-process, classifier, language model generation& N-grams, Markov model, k-means iterative clustering, Needleman-Wunsch string alignment for contextual dependency, TF/IDF, k-medoids algorithm \\ \hline

\texttt Prakash KDD'09 \cite{prakash2009bgp}& Automated BGP updates analyzer to find anomalies & Data flow, temporal analysis, frequency analysis& Haar wavelet transform, a frequency analysis technique akin to NB  \\ \hline

\texttt Qiu USENIX'09\cite{qiu2009locating:prefixhijacking}& Locate prefix hijackers by AS path& Target monitor, classifier, rank engine & Hierarchical clustering(metric: similarity of AS level paths to the prefix), rank by likelihood of seeing hijacking events \\ \hline

\texttt Song NDSS'09\cite{song2009:maliciousWebquery} & Detect anomalies in web traffic & Packet assembler, WL, feature extractor, classifier & N-grams feature extraction, Mixture-of-Markov-Chains model, K-means clustering \\ \hline

\texttt Antonakakis USENIX'10 \cite{antonakakis2010building:dynamicRepsys:DNS}& Dynamic reputation system for DNS& Honeypot, knowledge base, BL, reputation system classifier & Decision tree, logit boost, X-means clustering\\ \hline

\texttt Bilge NDSS'11 \cite{bilge2011exposure:j48dt}& Detect malicious DNS domains with passive DNS analysis & Traffic recorder, feature extractor, domain BL, learning module, classifier & C4.5 decision tree\\ \hline

\texttt Antonakakis USENIX'11 \cite{antonakakis2011detecting}& Detect malware domains at upper-level DNS hierarchy by passive DNS traffic analysis and global DNS query resolution & Knowledge base, learning module, training classifier, and feature extractor & RF \\ \hline

\texttt Fatemieh NDSS'11 \cite{fatemieh2011using:measurement:networkhost}& Protect integrity of measurement in whitespace networks& Traffic collector, signal propagation, seed trusted set& SVM(qudratic kernel) \\ \hline

\texttt Venkataraman NDSS'13\cite{venkataraman2013automatically}& Infer evolution of malicious activities in regions of Internet& Changes in prefixes and IP address distributions& Decision tree \\ \hline

\texttt Vissers NDSS'15 \cite{vissers2015parking:domainsRF}& Classify parked domains& Browser-based client side classifier& RF\\ \hline 

\end{tabular}
\end{table*}

\begin{table*}
\footnotesize
\caption{Password Meters \& Crackers, and CAPTCHA Breakers}
\begin{tabular}{|p{2.4cm}|p{7.6cm}|p{7cm}|} \hline
Study & Goal & Recognition or Classification Engine \\ \hline
\texttt Golle CCS'08\cite{golle2008machine:asirracaptcha}& Asirra CAPTCHA attack& SVM(RBF kernel) \\ \hline
\texttt Frank CCS'08\cite{frank2008prob:model:rbac}& Probabilistic model for role engineering in role-based control access(RBAC)& Disjoint Decomposition Model(a structural equivalent to infinite relational model), Gibbs sampling to infer DOM parameters, clustering \\ \hline
\texttt Zander Sec'08\cite{zander2008improved:samplinghiddenservice}& Measurement and sampling techniques to reveal hidden services& Remote clock-skew estimation and synchronized sampling \\ \hline
\texttt Weir SP'09\cite{weir2009password} & Password cracker using probabilistic context-free grammar& Context-free grammar with information about probability distribution of user passwords \\ \hline
\texttt Bursztein SP'11\cite{bursztein2011failure}& Attack noise-based non-continuous audio CAPTCHA & Regularized Least Squares Classification(RLSC) \\ \hline
\texttt Zheng CCS'11 \cite{Zheng2011:svmuser:mouse}& Behavioral metric for user verification by mouse movement& SVM(RBF kernel) \\ \hline
\texttt Brusztein CCS'11\cite{Bursztein2011:textbasedCAPTCHA:strongandweak}& Text-based CAPTCHA strength and weakness. Recommended engines in different phases of attacks, and principles for secured CAPTCHAs& SVM(RBF kernel), KNN \\ \hline 
\texttt Castelluccia NDSS'12\cite{castelluccia2012adaptive}& Password strength meter& Markov models built on n-gram database \\ \hline
\texttt Narayanan SP'12 \cite{narayanan2012feasibility:authorship}& Identify anonymous authors by stylometry on internet scale& RLSC, linear SVM, NB  \\ \hline
\texttt Dyer SP'12 \cite{dyer2012peek:sp:taattack}& Traffic analysis attack against common countermeasures& NB, multinomial NB, variable n-gram, SVM   \\ \hline
\texttt Kelley SP'12 \cite{kelley2012guess:passwordstrengh}& Measure password strength by simulating password-cracking algorithms& Brute force Markov models\\ \hline
\texttt Gao CCS'13 \cite{Gao:2013:attackHollowCaptcha}& Attack hollow textual CAPTCHAs& CNN  \\ \hline
\texttt Afroz SP'14 \cite{Afroz:stylometry_lr}& Detect multiple identities of anonymous authors& Principal component analysis(PCA), L1-regularized LR, linear SVM \\ \hline 
\texttt Ma SP'14\cite{ma2014study}& Study of probabilistic password models& Markov models and context-free grammar \\ \hline
\texttt Chatterjee SP'15 \cite{chatterjee2015cracking}& Natural language encoder design for cracking-resistant password vaults& Natural language encoder(NLE) \\ \hline
\texttt Freeman NDSS'16 \cite{freeman2016prob:userauthen}& Probabilistic model for user authentication at login time& Probabilistic model  \\ \hline
\texttt Gao NDSS'16\cite{gao2016simple}& Generic attack model on text-based CAPTCHAs& SVM(kernel unspecified), back-propagation neural network, template matching, CNN \\ \hline
\end{tabular}
\end{table*}

\begin{table*}
\footnotesize
\caption{Softward and Applications Security}
\begin{tabular}{|p{2.4cm}|p{3cm}|p{5.6cm}|p{6cm}|} \hline
Study & Goal & System Components & ML Techniques and Metrics(if specified) \\ \hline
\texttt Zhang Sec'08\cite{zhang2008highly}& Predict websites blacklisted by browsers& Security logs, pre-filter by knowledge base, relevance ranking, severity assessing, final BL & Link analysis and rank relevance correlation statistics on weighted undirected graph\\ \hline

\texttt Robertson NDSS'10\cite{robertson2010effective} & ML-based web app anomaly detection with scarce training data& Knowledge base(local and global), offline and online trained classifier & HMM-encoded probabilistic grammar, agglomerative hierarchical clustering \\ \hline

\texttt Chapman CCS'11\cite{Chapman2011:sidechannel:vulnerabilitydetector:webapp}& Quantify side channel leaks from web apps & Web app, web crawler, metrics and feature extraction, quantifier(classifier, entropy calculator, Fisher criterion calculator)&  Nearest-centroid(metrics: Total-Source-Destination, edit distance, random distance) \\ \hline

\texttt Sculley KDD'11\cite{sculley2011detectingad:adml}& Detect adversarial advertisements& Model aggregation, labeled data, stratified sampling, classifier, human monitoring feedback& Linear SVM and its variations \\ \hline

\texttt Curtsinger Sec'11\cite{Curtsinger:2011:malware}& In-browser JS malware detection & URL BL, scan scripts, code de-obfuscator, feature extractor, classifier& NB(metric: matched strings) \\ \hline

\texttt Lu CCS'11\cite{lu2011surf}& Browser component to detect malicious search poisoning& Browser/network/search/user redirection chain information collector, feature extractor, classifier& C4.5 decistion tree \\ \hline 

\texttt Kapravelos USENIX'13\cite{Kapravelos13:malware:KNN}& Detect evasions in malicious JS& Web information, drive-by-download detector oracle, syntax tree parser, similarity measurement, pairing, evasion detector& KNN(metric: direct editing distance between nodes)\\ \hline 

\texttt Bhagavatula AISec'14\cite{Bhagavatula2014:mlresourcefiltering}& Auto-block ad resources& Independent classifier(trained with labeled datasets)& KNN, experimented with NB, SVM(polynomial and RBF kernel, LR) \\ \hline

\texttt Soska Sec'14\cite{usenix2014badpageml}& Detect vulnerable websites before compromise& Websites' statistics and contents, parser and filter, dynamic feature extractor, classifier & C4.5 decision tree ensembles \\ \hline 

\texttt Borgolte Sec'15 \cite{usenix2015:imageobjrecog:webdeface}& Detect website defacement& URL of website(no prior knowledge of content or structure of webpages needed)& Stacked encoder, deep neural network, image-based object recognition \\ \hline 
\end{tabular}
\end{table*}

\begin{table*}
\footnotesize
\caption{Vulnerability and Security Policy Management}
\begin{tabular}{|p{2.4cm}|p{3cm}|p{4.5cm}|p{3.4cm}|p{3.7cm}|} \hline
Study & Goal & System Components & Similarity or Scoring Metrics& ML Techniques \\ \hline
\texttt Makanju KDD'09\cite{makanju2009clustering}& Mine clusters from event logs for fault management & 3-step partitions, independent system & Token- and bijection-based heuristics & Iterative hierarchical clustering   \\ \hline

\texttt Richardson AISec'10\cite{richardson2010limits}& Limits of automated fingerprinting for OS& Probe generator, candidate probe set, trainer, learner, feedback loop & Not applicable(NA)& C4.5 decision tree, rule learner, SVM-SMO, instance-based clustering \\ \hline

\texttt Bozorgi KDD'10\cite{bozorgi2010beyond} & Predict possibility and timing of vulnerable exploits& Vulnerability disclosure reports, feature extractor, offline classifier, online predictor& Common Vulnerabilities Scoring System(CVSS) score, distance to maximum margin hyperplane& Linear SVM\\ \hline

\texttt Peng CCS'12\cite{peng2012using:risksofandroid}& Score and rank risks of Android apps& App meta-information corpus, trainer, classifier& Heuristics derived from models& NB(basic and with prior), a hierarchical Bayesian model built as an extension to latent Dirichlet allocation(LDA)\\ \hline

\texttt Pandita USENIX'13\cite{WHYPER:usenix2013:nlp:spm}& Identify permission sentences in mobile apps& Semantic graph generator and engine, flow-of-logic engine, annotation engine, classifier(with user feedback loop to build a knowledge base) & NA& Part-of-speech(POS) tagging, phrase and clause parsing, named entity recognition(NER), semantic graph, typed dependency\\ \hline

\texttt Rasthofer NDSS'14\cite{rasthofer2014machine}& Identify sources and sinks from code of any Android API& Input, semantic and syntactic feature matrices, classifier& NA& Linear SVM\\ \hline

\texttt Yamaguchi SP'15\cite{yamaguchi2015automatic}& Infer vulnerability search patterns in C code& Static analyzer, signature generator, inference engine& Jaro distance& Complete-linkage clustering \\ \hline 

\texttt Wang USENIX'15\cite{usenix2015:semisup:EASEAndroid}& Semi-supervised learning for Android SPM& Knowledge base, classifier, co-occurrence learner, learning combiner, policy refinement generator feedback loop& Semantic pattern-to-rule distance& KNN(metric: semantic connectivity between known and unknown subjects)\\ \hline 

\texttt Liu USENIX'15 \cite{usenix2015:rfforecase:incident}& Predict future data leak instances from network logs& Data collection, BL, feature extractor, aggregation, training, classifier& Correlations between misconfigurations and malicious activities& RF \\ \hline 
\end{tabular}
\end{table*}

\begin{table*}
\footnotesize
\caption{Information Flow and DDoS}
\begin{tabular}{|p{2.4cm}|p{4cm}|p{5.5cm}|p{5.1cm}|} \hline
Study & Goal & System Components & ML Techniques  \\ \hline
\texttt Berra CCS'08(DDoS)\cite{berral2008adaptive:againstddos}& Detect and predict abnormal traffic patterns in distributed systems& Local model of information collector, classifier, learner& NB   \\ \hline

\texttt Venkataraman NDSS'08(DDoS)\cite{venkataraman2008limits}& Limits of pattern-extraction algorithms in adversarial setting& Adversary challenges the signature algorithm& Generic proof of bounded FPs and reflecting set of plausible aprioris\\ \hline

\texttt Yan CCS'12(DDoS)\cite{yan2012towards}& Framework to evaluate DDoS attacks and defense& Network state, attacker and defender's move spaces, reasonings, decisions& Game theory(semi network-form game), Bayesian network\\ \hline

\texttt Vsrndic NDSS'13(IF)\cite{vsrndic2013detection} & Static method to detect malicious PDF files& Hierarchical document structural feature extractor, classifier& C5.0 decision tree, SVM-RBF \\ \hline

\texttt Rndic SP'14(IF)\cite{rndic2014practical}& Evasion of learning-based PDF classifier& Attacker's knowledge of feature sets, training data, classification algorithms(one or more areas)& SVM(linear and RBF), RF as test classifiers \\ \hline

\texttt Lu NDSS'15(IF)\cite{lu2015checking}& Detect privacy leak in data flow& App information collector, data- and system-dependence graph, privacy disclosure analysis, peer voting engine& Ranking based on TF/IDF and cosine similarity representing semantic similarity \\ \hline

\texttt Xu NDSS'16(IF)\cite{xu2016automatically}& Generic automated evasions of malicious PDF classifiers& Population initialization, mutation, variant selection, feedback to population& SVM-RBF, RF  \\ \hline

\end{tabular}
\end{table*}

\begin{table*}
\footnotesize
\caption{Malware Identification, Classification, and Mitigation}
\begin{tabular}{|p{2.4cm}|p{4cm}|p{5.5cm}|p{5.1cm}|} \hline
Study & Goal & System Components& ML Techniques \\ \hline
\texttt Ahmed CCS'09\cite{ahmed2009using:spatialtemporal:malware}& Detect malware by Windows API call traces & Kernel mode hook, knowledge base, API call trace, feature extractor, training, classifier& C4.5 decision tree, instance-based KNN, NB, inductive rule learner, SVM-SMO\\ \hline

\texttt Bayer NDSS'09\cite{bayer2009scalable}& Group similar malwares from taint source and control flow& Knowledge base, dynamic analysis, network analysis, behavior profile generator, clustering& Locality sensitive hashing(LSH), single linkage clustering. Metric: Jaccard distance, normalized compression distance \\ \hline

\texttt Comparetti SP'09\cite{Comparetti2009:prospex}& Automated reverse engineering of app layer protocol specifications & Dynamic analysis, session analysis, message clustering, state machine inference, labeling, fuzzing& Partitioning around medoids(PAM), metric: Jaccard index \\ \hline

\texttt Kolbitsch Sec'09\cite{Kolbitsch2009:malware:dag:graphmatching}& Detect malware at end hosts& Knowledge base, program slicing, behavioral profile generation, matching& Directed acyclic graph(DAG) matching algorithms \\ \hline

\texttt Lanzi CCS'10\cite{Lanzi2010:accessminer:malware:ngram}& Diversity of system calls study(System-centric malware analysis)& Training dataset, system call sequence miner& N-gram models \\ \hline

\texttt Ye KDD'10\cite{ye2010automatic}& Auto-group malware samples into families with a cluster ensemble& Feature extractor, n-gram slicer, cluster ensemble, signature generator, human feedback loop& PCA, Ensemble: hierarchical \& partitional clustering, with weighted subspace K-medoids \\ \hline

\texttt Fredrikson SP'10 \cite{Fredrikson:2010malwaresmanl}& Auto-extraction on specifications of class of programs & Training datasets, behavior miner, extracted specifications& Simulated annealing in graph algorithm \\ \hline 

\texttt Jang CCS'11\cite{Jang2011:featurehashing:malwaresemantics:bitshred}& Reduce feature dimensionality for large-scale malware clustering& Static and dynamic analysis, fingerpriting, scheduler, clustering engine & Feature hashing(metric: Jaccard similarity), aggolomerative hierarchical clustering\\ \hline 

\texttt Ye KDD'11\cite{ye2011combining}& Detect malware by file relations& File relation and content collector, feature extractor, classifier & Customized parametric model on content and non-parametric one on relations  \\ \hline

\texttt Antonakakis Sec'12\cite{usenix2012:clustermalware}& Detect domain generation algorithm(DGA)-based malwares& Knowledge base, discovery engine, trainer, classifier within networks& X-means clustering, spectral clustering, decision tree, hidden Markov model(HMM) \\ \hline 

\texttt Rahman Sec'12\cite{usenix2012:svm:osn:socware}& Detect malwares propagated by social network& User authorization, post crawler, feature extractor, WL, BL, train(manually labeled data) classifier, user feedback& SVM(kernel unspecified) \\ \hline 

\texttt Kong KDD'13\cite{kong2013discriminant:metriclearning}& Malware classification on structural information& Labeled set, function call graph(FCG) extractor, maximum margin distance learner, ensemble classifier& SVM-Gaussian kernel, KNN (metric: maximum margin-guided similarity between FCGs)\\ \hline

\texttt Borgolte CCS'13\cite{borgolte2013delta}& Signature-based malware clustering& Website parser, DOM tree difference computation engine, feature extractor, signature generator&  Density-based clustering with ordering points(metrics: Shannon entropy, Kolmogorov complexity) \\ \hline

\texttt Biggio AISec'14\cite{biggio2014poisoning}& Perfect-knowledge attacker poisoning malware clustering& Malware instruction set representation, message embedder, clustering, classifier& Single linkage clustering   \\ \hline

\texttt Tamersoy KDD'14\cite{tamersoy2014guilt}& Detect malware by file relation graphs& File collector, LSH, graph builder, belief propagator& MinHash, LSH, pairwise Markov random field, unweighted bipartite graph   \\ \hline

\texttt Invernizzi NDSS'14\cite{invernizzi2014nazca}& Detect malware downloads in networks& Network traffic collector, feature extractor, distributed classifier& Decision tree(ground truth)  \\ \hline 

\texttt Arp NDSS'14\cite{arp2014drebin}& Explainable Android malware detection& Broad static analysis, feature embedder, detector, explanation& Linear SVM   \\ \hline

\texttt Graziano Sec'15 \cite{usenix2015:lmt:malwaredev}& Detect and forecast malware samples and trends from public dynamic analysis sandbox& Dynamic analysis, binary similarity, fine-grained static analysis, classifier& Logistic model tree  \\ \hline

\texttt Shin Sec'15\cite{usenix2015:rnn:binary:function}& Recognize functions in malware binaries& Binary, fixed-length subsequence extractor, learner& Recurrent neural network(RNN) with 1 hidden layer(optimized with stochastic gradient descent ``SGD'')  \\ \hline

\texttt Kirat CCS'15\cite{kirat2015malgene:idf:longestsubsequence}& Auto-extract malware evasion behavior signature& Execution event extractor, call sequence alignment, event comparison, clustering& Local sequence alignment, IDF, hierarchical clustering(metric: Jaccard similarity) \\ \hline

\texttt Smutz NDSS'16 \cite{smutz2016tree}& Detect malware mimicry evasions with ensemble classifiers& Two malware classifiers as an ensemble, using mutual agreement analysis& Ensemble: Linear SVM, RF  \\ \hline
\end{tabular}
\end{table*}

\begin{table*} 
\footnotesize
\caption{Phishing, Malicious Content and Behaviors}
\begin{tabular}{|p{2.4cm}|p{3cm}|p{3.2cm}|p{4.9cm}|p{3.5cm}|} \hline
Study & Goal & Features & System Components & ML Techniques \\ \hline
\texttt Chatterjee AISec'08\cite{chatterjee2008robust:repsys}& Content-based reputation system robust against Sybil attacks& Content evolution, contribution history& Editing history, local and global reputation growth and correlation, scoring& Customized reputation scoring functions\\ \hline

\texttt Ma ICML'09\cite{ma2009identifyingURL}& Detect malicious websites by URLs in active learning& Lexical- and host-based features, no content features & Live URL feed, labeling engine, feature extractor, classifier with feedback loop& Perceptron, LR(with SGD), passive-aggressive and confidence-weighted algorithm \\ \hline

\texttt Ma KDD'09\cite{ma2009beyondURL:kdd}& Detect malicious websites from URLs& Lexical- and host-based features of URLs & Complementary to BLs & NB, SVM-RBF, linear SVM, L1-regularized LR \\ \hline

\texttt Hao Sec'09\cite{hao2009detecting:reputationsys}& Automated reputation engine for spam filter& Network-level and SMTP header features, temporal patterns & Emails, BL, WL, feature extractor, classifier, policy pool & A rule-based linear learning ensemble \\ \hline

\texttt Blum AISec'10\cite{Blum2010:phishingurl:lexicalfeatures}& Detect existing and emerging phishing domains & Lexical- and content-based features& Rule-based BL, online learning classifier& Deep MD5 matching(metric: Kulczynski 2 coefficient) \\ \hline

\texttt Whittaker NDSS'10\cite{whittaker2010large}& Detect phishing web pages & Hosting properties, content-based and URL-structure features& Search engine BL, URL aggregator, feature extractor, classifier& LR - online gradient descent \\ \hline

\texttt Thomas SP'11\cite{thomas2011design}& Real-time URL spam filter& URL lexical, content, hosting property, browser, DNS resolver, IP analysis features& Web service with URL stream, URL aggregator, feature collector and extractor, classifier, feedback loop(BL annotation training)& L1-regularized LR, LR-SGD \\ \hline

\texttt Br\"{u}ckner KDD'11\cite{Bruckner:2011kdd:spamfilter}& Stackelberg game model in spam filtering& A data generator reacts to the learner's moves& Learner, data generator, loss function& Stackelberg-based prediction game with different loss functions\\ \hline

\texttt Lee NDSS'12\cite{lee2012warningbird}& Detect malicious URLs in streams& Properties of redirect chains of URLs& URL stream data collection, feature extraction, training & L2-regularized LR\\ \hline

\texttt Afroz SP'12 \cite{afroz2012detecting:writingstyle}& Detect frauds in online writing styles & Lexical-, syntactic-, content-specific features& Feature extractor, classifier & SVM-SMO, experimented with NB, SVM-RBF, C4.5, LR \\ \hline 

\texttt Egele NDSS'13\cite{egele2013compa:ndssnlp}& Detect compromised accounts by behavior change& Temporal behavior and content patterns& Data collection, labeling, training& N-gram model, SVM-SMO\\ \hline

\texttt Gro{\ss}hans ICML'13\cite{grosshans2013bayesian}& Bayesian game model in adversarial spam filtering& Game between a learner of a model and a data generator& Learner, data generator(conflicting but not necessarily adversarial), cost functions& Bayesian regression model. Baseline: Nash regression, robust ridge regression, regular ridge regression\\ \hline

\texttt Invernizzi SP'13 \cite{invernizzi2012evilseed}& Use search engines to find other malicious webpages& Content- and link-based features& Crawler, profiler, search engine's BL, initial set of malicious pages& N-grams, term extraction   \\ \hline

\texttt Zhao KDD'13\cite{zhao2013cost:malurl:2013kdd}& Online active learning for malicious URL detection& Lexical- and host-based features& Live data feed, feature collector, cost-sensitive update, active learning module, classifier& Customized online active cost-sensitive algorithm \\ \hline 

\texttt Jiang Sec'13 \cite{usenix2013:bayesian:spam:phonenumber}& Detect spam cell numbers& Call detail records& Records collector, feature extractor, classifier& Customized Bayesian model\\ \hline

\texttt Zhang NDSS'14\cite{zhang2014dspin:measurement:nlp}& Detect auto-generated web spam content& Content features& Content filter, inverted index engine, clustering& Shingling, POS tagging(metric:Jaccard index) \\ \hline

\texttt Viswanath Sec'14\cite{usenix2014:anomuserbehavior}& Unsupervised anomalous user behavior detection& Spatial, temporal, spatial-temporal& Normality construction, PCA, detection& PCA, KNN \\ \hline

\texttt Whalen AISec'14\cite{whalen2014model}& Distributed content anomaly detection(CAD)& N-gram of payloads& Distributed models over application servers& Aggregated RF, LR, Bloom filter\\ \hline

\texttt Boshmaf NDSS'15\cite{boshmaf2015integro}& Predict benign users who befriended fake accounts& User profiles, communications, activities history& Graph representation, feature extraction, trust propagation, rank users& Louvain method for community detection, modified random walk\\ \hline

\texttt Stringhini Sec'15\cite{usenix2015:ml:malicioususer}& Detect malicious online accounts accessed by a common set of machines& User-agent correlation, account usage, event time series& IP mapping, graph representation, feature extraction, detection& Louvain method\\ \hline

\texttt Wang Sec'15\cite{usenix2015:adml:crowdsourcer:svmrf}& Detect adversarial crowdsourcing& User profile, behaviors, temporal patterns& Normality construction, input, feature extraction, classification& Customized Bayesian model, SVM(RBF and polynomial), RF \\ \hline
\end{tabular}
\end{table*}

\begin{table*} \footnotesize
\caption{Intrusion Detection System Research}
\begin{tabular}{|p{2.4cm}|p{4cm}|p{5.8cm}|p{4.7cm}|} \hline
Study & Goal & Key Assumption(s) & ML Techniques \\ \hline
\texttt Kloft AISec'09\cite{Kloft2008:autofeaturesselection}& Auto-select features for anomaly detection with one-class SVM& Different feature sets might be various characterizations of normality & Generalized support vector data description(SVDD) \\ \hline

\texttt Gornitz AISec'09\cite{Gornitz:2009:activeLearningIDS}& Query low-confidence observations and expand data basis with minimal labeling& Anomaly detection is an active learning task& SVDD \\ \hline

\texttt Symons AISec'12 \cite{symons2012nonparametric}& Non-parametric semi-supervised network IDS. & Attackers' behaviors are not necessarily different from legitimate human users, e.g. they do not sent large numbers of probes & Laplacian regularized least squares(RLS)/Bayesian kernel model on graphs representing network flow traffic\\ \hline

\texttt Ding KDD'12 \cite{ding2012intrusion:measurement:ids} & Augment signature-based system at network edges for multi-layer IDS & Attackers do not respect community boundaries and exhibit ``anti-social behaviors''& Graph-based community detection \\ \hline

\texttt Xie KDD'12 \cite{xie2012review}& Detect spam review by temporal patterns in adversarial environment& Normal reviewers' arrival patterns are Poisson and uncorrelated to rating patterns temporally& Multidimensional time series(matching blocks by longest common substring algorithm) \\ \hline

\texttt Wressnegger CCS'13 \cite{wressnegger2013close}& Use of n-grams in intrusion detection& Anomaly detection and ML are two different schemes for IDS& Reviewed n-gram models in both ML and anomaly detection schemes \\ \hline

\texttt Momtazpour KDD'15\cite{momtazpour2015analyzing}& Detect correlated invariants in cyber-physical systems & Local invariants after filtering can still be correlated & Latent factor auto regression with exogenous input to correlate time series \\ \hline

\texttt Juba NDSS'15\cite{juba2015principled}& Reproducibility and comparison framework for IDS& Uniform distribution approximates the entire web by Common Crawl URL index& Probablistic model\\ \hline 
\end{tabular}
\end{table*}

\bibliographystyle{IEEEtran}

\end{document}